\documentclass[10pt,journal,onecolumn]{IEEEtran}

\usepackage{amsfonts}
\usepackage{amsmath}
\usepackage{amssymb}
\usepackage{bm}
\usepackage{graphicx}
\usepackage{color, soul}
\usepackage{stfloats}
\usepackage[numbers,sort&compress]{natbib}
\usepackage[amsmath,thmmarks]{ntheorem}
\usepackage{theorem}
\usepackage{subfigure}
\usepackage{ntheorem}
\usepackage{url}
\usepackage{hyperref}
\usepackage{float}
\usepackage{mathrsfs}
\usepackage{algorithm}
\usepackage{algorithmic}
\usepackage{mathrsfs}
\usepackage{algorithm}
\usepackage{algorithmic}

\newcommand{\ua}{\uparrow}
\usepackage{color, soul}
\newcommand{\nc}{\newcommand}
\nc{\da}{\downarrow} \nc{\hc}{\hat{c}} \nc{\hS}{\hat{S}}
\nc{\bra}{\langle} \nc{\ket}{\rangle} \nc{\eq}{equation (\ref}
\nc{\h}{\hat} \nc{\hT}{\h{T}}\nc{\be}{\begin{eqnarray}}
\nc{\ee}{\end{eqnarray}}\nc{\rd}{\textrm{d}}\nc{\e}{eqnarray}\nc{\hR}{\hat{R}}\nc{\Tr}{\mathrm{Tr}}
\nc{\tS}{\tilde{S}}\nc{\tr}{\mathrm{tr}}\nc{\8}{\infty}\nc{\lgs}{\bra\ua,\phi|}\nc{\rgs}{|\ua,\phi\ket}
\nc{\hU}{\hat{U}}\nc{\lfs}{\bra\phi|}\nc{\rfs}{|\phi\ket}\nc{\hZ}{\hat{Z}}\nc{\hd}{\hat{d}}\nc{\mD}{\mathcal{D}}
\nc{\bd}{\bar{d}}\nc{\bc}{\bar{c}}\nc{\mc}{\mathcal}\nc{\ea}{eqnarray}\nc{\mG}{\mathcal{G}}\nc{\bce}{\begin{center}}
\nc{\ece}{\end{center}}

\theoremheaderfont{\sc}\theorembodyfont{\upshape}
\theoremstyle{nonumberplain}
\theoremseparator{}
\theoremsymbol{\rule{1ex}{1ex}}

\usepackage{indentfirst}
\par
\usepackage{graphics}
\IEEEoverridecommandlockouts

\begin{document}

\title{Binary Sparse Bayesian Learning Algorithm for One-bit Compressed Sensing}

\author{Jiang Zhu, Lin Han, Xiangming Meng and Zhiwei Xu
}
%
%
%
%

\maketitle

\begin{abstract}
In this letter, a binary sparse Bayesian learning (BSBL) algorithm is proposed to slove the one-bit compressed sensing (CS) problem in both single measurement vector (SMV) and multiple measurement vectors (MMVs). By utilising the Bussgang-like decomposition, the one-bit CS problem can be approximated as a standard linear model. Consequently, the standard SBL algorithm can be naturally incorporated. Numerical results demonstrate the effectiveness of the BSBL algorithm.
\end{abstract}

\section{Introduction}
{Compressed sensing (CS) aims to reconstruct sparse signals from the underdetermined measurements \cite{Donoho1}, which has many applications in Magnetic Resonance Imaging (MRI), lensless imaging and network tomography \cite{Donoho2, Huang, Firooz}. Various algorithms have been proposed to solve the standard linear models (SLMs) , such as the least absolute shrinkage and selection operator (lasso) \cite{Robert}, the iterative thresholding algorithms \cite{Daubechies}, the sparse Bayesian learning (SBL) algorithm \cite{Tipping, WipfS} and approximate message passing (AMP) algorithms \cite{Donoho}.}

With the rapid development of the millimeter wave (mmWave) communication technology, the future communication transmission rate will be greatly improved, which means that the sampling rate of the analog-to-digital converter (ADC) must be increased. However, high-speed high-precision ADC is either not available, or is costly and power-hungry \cite{Singh}. One approach to reduce power consumption is to adopt the low-precision quantized systems (1-4 bits). {In this setting, traditional algorithms for SLMs suffer performance degradation in the low precision quantized systems.} As a result, it's of great theoretical and practical importance to study the channel estimation and direction of arrival (DOA) from quantized measurements, especially for the one-bit scenarios.

One of the representative algorithms for solving the one-bit CS reconstruction problems is the binary iterative hard thresholding (BIHT) algorithm, which was proposed to recover the original signals from noiseless one-bit measurements and having a remarkable performance in terms of the reconstruction error as well as consistency \cite{BIHT}.  For the  generalized linear models (GLMs), a generalized approximate message passing (GAMP) algorithm was proposed \cite{Rangan1}. It is shown that the GAMP algorithm can be applied to solve the channel estimation problems for mmWave multiple input multiple output (MIMO) systems with one-bit ADCs \cite{Mo}. In addition, the GAMP algorithm can be naturally incorporated into the adaptive quantization framework \cite{Kamilov,Hangting}.  In \cite{meng1}, the SBL algorithm is extended to deal with the GLMs, where the GLMs are iteratively approximated as SLMs, and a generalized SBL (Gr-SBL) algorithm is proposed. Besides, the Gr-SBL is applied to solve the one-bit DOA estimation problem \cite{meng2}. The Gr-SBL algorithm leverages the joint sparsity of the real and imaginary parts and thus improves the recovery performance.

In this paper, utilising the Bussgang-like decomposition \cite{Bussgang}, we transfer a nonlinear model into a linear one. Then we naturally propose the binary sparse Bayesian learning (BSBL) algorithm.  Compared to the Gr-SBL algorithm \cite{meng1, meng2}, which approximates the nonlinear model into the pseudo linear one with iteratively updated measurements, the binary observation of the BSBL algorithm is kept unchanged during the whole iteration process. The simulation results have shown its effectiveness in the single measurement vector (SMV) and multiple measurement vectors (MMVs). In addition, this BSBL algorithm can also be applied to the correlated noise scenario.

Notation: For a vector ${\mathbf x}$, let ${\rm diag}({\mathbf x})$ denote a matrix whose diagonal is composed of $\mathbf x$. For a square matrix ${\mathbf A}$, let ${\rm diag}({\mathbf A})$ denote a vector whose elements are the diagonal elements of ${\mathbf A}$. For a positive definite matrix ${\mathbf S}$, let ${\rm diag}({\mathbf S})^{-\frac{1}{2}}$ denote the elementwise inverse square root of ${\rm diag}(S)$. For the scalar $c$ and vector ${\mathbf a}$, the division operation ${c}/{\mathbf a}$ and power operation ${\mathbf a}^2$ are applied componentwisely.

\section{Algorithm}
In this section, the BSBL algorithm for one-bit CS in both SMV and MMVs scenarios are proposed. The key step is to iteratively transform the one-bit quantization problem into a linear model, then the SBL algorithm can be naturally incorporated.

\subsection{{Single measurement vector}}
Consider the estimation problem from one-bit measurements described as (extended to complex observations)
\begin{align}\label{singlesnap}
{\mathbf y} = {\rm {csgn}}\left({\mathbf A}{\mathbf x}+{\mathbf w}\right),
\end{align}
where ${\rm csgn}({\mathbf r})={\rm sgn}({\rm Re}({\mathbf r}))+{\rm sgn}({\rm Im}({\mathbf r})){\rm j}$, ${\rm Re}({\mathbf r})$ and ${\rm Im}({\mathbf r})$ denote the real and imaginary parts of ${\mathbf r}$, ${\rm sgn}(\cdot)$ returns the componentwise sign of its variables. ${\mathbf y}\in {\mathbb C}^M$ denotes the $M$ complex binary-valued measurements. $\mathbf A\in {\mathbb {C}}^{M\times N}$ is a known measurement matrix, ${\mathbf w}\sim {\mathcal {CN}}({\mathbf w};{\mathbf 0},{\mathbf C}_{\mathbf w})$ is a circular symmetric Gaussian noise with covariance matrix being ${\mathbf C}_{\mathbf w}$. $\mathbf x \in{\mathbb {C}}^{N\times 1}$ denotes the complex amplitudes, whose number of non-zero elements is $K$. Equivalently, the complex observation model (\ref{singlesnap}) can be equivalently expressed as
\begin{align}\label{realsinglesnap}
\bar{\mathbf y} = {\rm {sgn}}\left(\bar{\mathbf A}\bar{\mathbf x}+\bar{\mathbf w}\right),
\end{align}
where $\bar{\mathbf w}\sim {\mathcal {N}}(\bar{\mathbf w};{\mathbf 0},{\mathbf C_{\bar{\mathbf w}}})$ and
\begin{align}
\bar{\mathbf y} =\left[\begin{array}{c}
                   {\rm Re}({\mathbf y}) \\
                   {\rm Im}({\mathbf y})
                 \end{array}\right],
\bar{\mathbf A} = \left[\begin{array}{cc}
                    {\rm Re}({\mathbf A}) & -{\rm Im}({\mathbf A}) \\
                    {\rm Im}({\mathbf A}) & {\rm Re}({\mathbf A})
                  \end{array}\right],\\
\bar{\mathbf x} =\left[\begin{array}{c}
                   {\rm Re}({\mathbf x}) \\
                   {\rm Im}({\mathbf x})
                 \end{array}\right],
{\mathbf C_{\bar{\mathbf w}}} = \frac{1}{2}\left[\begin{array}{cc}
                    {\rm Re}({\mathbf C}_{\mathbf w}) & -{\rm Im}({\mathbf C}_{\mathbf w}) \\
                    {\rm Im}({\mathbf C}_{\mathbf w}) & {\rm Re}({\mathbf C}_{\mathbf w})
                  \end{array}\right].
\end{align}

In the following, we focus on solving the problem (\ref{realsinglesnap}) instead of (\ref{singlesnap}), as (\ref{realsinglesnap}) can be transformed as a linear regression problem \cite{Andrew}. For the SBL framework, the priors of elements of $\bar{\mathbf x}$ are independent and identically distributed (i.i.d.) Gaussian random variables \cite{SBL},\footnote{One can see that for complex signals, its real and imaginary parts should share some common sparsity. While in this paper, we assume that they are independent and some performance loss may be incurred due to this assumptions.} i.e.,
\begin{align}\label{xx}
p(\bar{x}_i|\alpha_i)=\sqrt{\frac{\alpha_i}{2\pi}}{\rm e}^{-\alpha_i\bar{x}_i^2/2}\triangleq {\mathcal {N}}(\bar{x}_i;0,\alpha_i^{-1}),
\end{align}
where ${\boldsymbol \alpha}=[\alpha_1,\cdots,\alpha_{2N}]^{\rm T}$ contains the hyperparameters that control the sparsity of $\bar{\mathbf x}$, $(\cdot)^{\rm T}$ denotes the transpose operator and {${\mathcal {N}}(\bar{x}_i;0,\alpha_i^{-1})$ denotes the Gaussian probability density function (PDF) of $\bar{x}_i$ with mean 0 and variance $\alpha_i^{-1}$}. Assuming that each element of ${\boldsymbol \alpha}$ follows the Gamma distribution expressed as
\begin{align}\label{alphaalpha}
p(\alpha_i)={\rm Ga}(\alpha_i|a,b)=\frac{b^a}{\Gamma(a)}\alpha_i^{a-1}{\rm e}^{-\alpha_ib},
\end{align}
here ${\rm Ga}(\alpha_i|a,b)$ denotes the Gamma distribution with shape $a$ and rate $b$, and ${\Gamma(a)}$ is the Gpamma function
\begin{align}
{\Gamma(a)}\triangleq \int_{0}^{\infty}u^{a-1}{\rm e}^{-u}{\rm d}u.
\end{align}
Note that $a=1$ and $b=0$ corresponds to the uninformative prior of ${\boldsymbol \alpha}$.

Now the BSBL algorithm is described in detail. For the $t$-th iteration, assume that $\bar{\mathbf x}\sim {\mathcal {N}}(\bar{\mathbf x};{\mathbf 0},{\mathbf C}_{\bar{\mathbf x}}(t))$, where ${\mathbf C}_{\bar{\mathbf x}}(t)\triangleq {\rm diag}({\boldsymbol \alpha}^{-1}(t))$ and $(\cdot)^{-1}$ denotes the componentwise division. By utilising the Bussgang-like decomposition, model (\ref{realsinglesnap}) can be approximated as a linear regression problem \cite{Andrew}
\begin{align}\label{dic}
\overline{\mathbf y}={\mathbf F}(t)\bar{\mathbf x}+{\mathbf e}(t),
\end{align}
where ${\mathbf F}(t)\in {\mathbb {R}}^{2M\times 2N}$ is a \emph{linearization matrix} and ${\mathbf e}(t)\in {\mathbb {R}}^{2M\times 1}$ is a \emph{residual error vector} that includes noise and linearization artifacts. The closed-form expression of $\mathbf F$ can be easily obtained \cite{Andrew}.

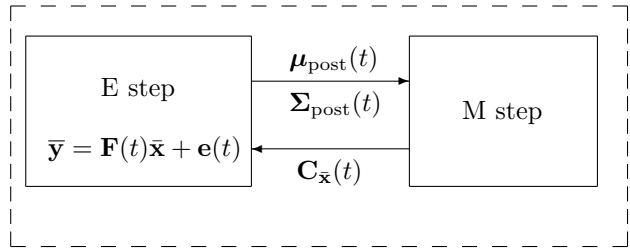
\begin{figure}[h!t]
\centering
\setlength{\unitlength}{1cm}
\begin{picture}(14,4)
\multiput(4,0.8)(0.4,0){21}{\line(1,0){0.2}}
\multiput(4,4.0)(0.4,0){21}{\line(1,0){0.2}}

\multiput(4,1)(0,0.3){11}{\line(0,-1){0.2}}
\multiput(12.2,1)(0,0.3){11}{\line(0,-1){0.2}}

\put(4.2,1.6){\line(1,0){3}}
\put(4.2,3.6){\line(1,0){3}}
\put(4.2,3.6){\line(0,-1){2}}
\put(7.2,1.6){\line(0,1){2}}
\put(5.2,2.8){$\rm E~ step$}
\put(4.5,2.0){$\overline{\mathbf y}={\mathbf F}(t)\bar{\mathbf x}+{\mathbf e}(t)$}

\put(7.2,3.0){\vector(1,0){2.1}}
\put(9.3,2.1){\vector(-1,0){2.1}}
\put(7.7,3.2){${\boldsymbol \mu}_{\rm post}(t)$}
\put(7.7,2.6){${\boldsymbol \Sigma}_{\rm post}(t)$}
\put(7.8,1.7){${\mathbf C}_{\bar{\mathbf x}}(t) $}

\put(9.3,1.6){\line(1,0){2.5}}
\put(9.3,3.6){\line(1,0){2.5}}
\put(9.3,3.6){\line(0,-1){2}}
\put(11.8,1.6){\line(0,1){2}}
\put(10.0,2.5){${\rm M~step}$}
\end{picture}
\caption{System diagram for performing the BSBL algorithm}
\label{fig_model}
\end{figure}

{We propose an BSBL algorithm, which consists of two steps, as shown in Fig. \ref{fig_model}.}
{According to E step, we obtain the posterior means ${\boldsymbol \mu}_{\rm post}(t)$ and covariance matrix ${\boldsymbol \Sigma}_{\rm post}(t)$ of $\hat{\mathbf x}$ and pass them as the input of the M step. M step subsequently utilises the information of ${\boldsymbol \mu}_{\rm post}(t)$ and ${\boldsymbol \Sigma}_{\rm post}(t)$ to provide ${\mathbf C}_{\bar{\mathbf x}}(t) $ as the input of the E step. Then ${\mathbf C}_{\bar{\mathbf x}}(t) $ is used to update the model (\ref{dic}). This procedure proceeds until the number of iterations is reached. Now we present the details as follows.} We are interested in evaluating the E step, i.e., obtaining the posterior means and covariance matrix of $\bar{\mathbf x}$. According to \cite{Andrew, Li}, the posterior means and covariance matrix of $\bar{\mathbf x}$ are
\begin{subequations}\label{post_x}
\begin{align}
&{\boldsymbol \mu}_{\rm post}(t)={\mathbf E}^{\rm T}{\mathbf C}_{\overline{\rm y}}^{-1}\overline{\mathbf y},\label{postmeansingle}\\
&{\boldsymbol \Sigma}_{\rm post}(t)={\mathbf C}_{\bar{\mathbf x}}(t)-{\mathbf E}^{\rm T}{\mathbf C}_{\overline{\rm y}}^{-1}{\mathbf E},\label{postvarsingle}
\end{align}
\end{subequations}
where
\begin{subequations}\label{4}
\begin{align}
&{\mathbf C}_{\bar{\mathbf z}}= \bar{\mathbf A}{\mathbf C}_{\bar{\mathbf x}}(t) {\bar{\mathbf A}}^{\rm T} + {\mathbf C_{\bar{\mathbf w}}},\\
&{\mathbf E} = \left(\frac{2}{\pi}\right)^{1/2}{\rm {diag}}\left({\rm {diag}}({\mathbf C}_{\bar{\mathbf z}})^{-1/2}\right) \bar{\mathbf A}{\mathbf C}_{\bar{\mathbf x}}(t)\label{4c},\\
&{\mathbf C}_{\overline{\mathbf y}} = \left(\frac{2}{\pi}\right){\rm {arcsin}}\left({\rm {diag}}\left({\rm {diag}}({\mathbf C}_{\bar{\mathbf z}})^{-1/2}\right) \right)\left({\mathbf C}_{\bar{\mathbf z}} {\rm {diag}}\left({\rm {diag}}({\mathbf C}_{\bar{\mathbf z}})^{-1/2}\right)\right)\label{4d}.
\end{align}
\end{subequations}
To make the algorithm more stable, we propose using a damping factor on the posterior means and variances as
\begin{subequations}\label{damp}
\begin{align}
&\bar{\boldsymbol \mu}_{\rm post}(t) = \gamma{\boldsymbol \mu}_{\rm post}(t) + (1-\gamma)\bar{\boldsymbol \mu}_{\rm post}(t-1)\label{mean},\\
&\bar{\boldsymbol \Sigma}_{\rm post}(t) = \gamma{\boldsymbol \Sigma}_{\rm post}(t) + (1-\gamma)\bar{\boldsymbol \Sigma}_{\rm post}(t-1)\label{variance}.
\end{align}
\end{subequations}
For the M step, we update ${\boldsymbol \alpha}$ as
\begin{align}\label{alpha_update}
{\boldsymbol \alpha}(t+1)=\frac{2a-1}{2b+{\rm diag}(\bar{\boldsymbol \Sigma}_{\rm post}(t))+(\bar{\boldsymbol \mu}_{\rm post}(t))^2}.
\end{align}
Now, we close the loop of the BSBL algorithm. The proposed algorithm is summarized in Algorithm \ref{algorithmsingle}.
\begin{algorithm}[h]
\caption{BSBL algorithm for one-bit compressed sensing with SMV}
\begin{algorithmic}[1]\label{algorithmsingle}
\STATE Initialize ${\boldsymbol \alpha}(1)$, $\bar{\boldsymbol \mu}_{\rm post}(0)$ and $\bar{\boldsymbol \Sigma}_{\rm post}(0)$; Set the parameters of the noise covariance matrix {${\mathbf C}_{\bar{\mathbf w}}$}, the number of iterations $T$ and the damping factor $\gamma$;
\FOR {$t=1,\cdots,T$ }
\STATE Perform the E step and calculate the post means and variances of $\bar{\mathbf x}$ as (\ref{post_x}).
\STATE Perform the damping step (\ref{damp}).
\STATE Perform the M step and update ${\boldsymbol \alpha}$ as ${\boldsymbol \alpha}(t+1)$ (\ref{alpha_update}).
\ENDFOR
\STATE Return $\hat{\mathbf x}$.
\end{algorithmic}
\end{algorithm}

In some practical one-bit CS scenarios, the noise variance may be unknown and needs to be estimated. But when the noise is uncorrelated, i.e., $\mathbf C_{\mathbf w} = \sigma_w^2{\mathbf I}_M$, where ${\mathbf I}_M$ denotes the identity matrix with dimension $M$, and the relative amplitude or the support of $\mathbf x$ is the focus, there's no need to estimate $\sigma_w^2$ for Algorithm \ref{algorithmsingle} if we use the uninformative prior. In fact, the DOAs are determined by the relative amplitudes, not the complex amplitudes. Given that the noise variance is unknown, we can reformulate the model (\ref{singlesnap}) as
\begin{align}
{\mathbf y} = {\rm {csgn}}\left({\mathbf A}{\mathbf x}/\sigma_w+{\mathbf w}/\sigma_w\right).
\end{align}
In this setting, the variance of the additive noise ${\mathbf w}/\sigma_w$ is unit. The above analysis also applies to the Gr-SBL method, which is also explained in \cite{meng2}. In the numerical simulations, we will verify this fact.

\subsection{{Multiple measurement vectors}}
For the MMVs scenario, the model can be expressed as
\begin{align}\label{Multisnap}
{\mathbf Y}={\rm csgn}({\mathbf A}{\mathbf X}+{\mathbf W}),
\end{align}
where ${\mathbf Y}=[{\mathbf y}_1,\cdots,{\mathbf y}_L]\in {\mathbb C}^{M\times L}$ denotes the one-bit quantized measurements, $L$ is the number of snapshots and $M$ is the number of measurements per snapshot. {For each snapshot, the noise ${\mathbf w}_l$ satisfies ${\mathbf w}_{l}\sim {\mathcal {CN}}({\mathbf w}_{l};{\mathbf 0}, {\mathbf C}_{\mathbf w})$ and is independent across all the snapshots.} ${\mathbf X}=[{\mathbf x}_1,\cdots,{\mathbf x}_L]\in {\mathbb C}^{N\times L}$ is row sparse. The MMVs model can be decoupled as
\begin{align}
{\mathbf y}_l={\rm csgn}({\mathbf A}{\mathbf x}_l+{\mathbf w}_l),\quad l=1,\cdots,L.
\end{align}
Therefore, we apply Algorithm \ref{algorithmsingle} for each snapshot. In detail, for a given ${\boldsymbol \alpha}(t)$, we obtain the posterior means and covariance matrix of $\bar{\mathbf x}_l$ for each snapshot according to (\ref{post_x}). Note that the covariance matrix is the same for all the snapshots. Let $\bar{\boldsymbol \mu}_{{\rm post},l}(t)$ and $\bar{\boldsymbol \Sigma}_{\rm post}(t)$ denote the damped poster means and covariance matrix for the $l$-th snapshot. We now evaluate the expected complete data loglikelihood function $Q({\boldsymbol \alpha})$ given by
\begin{align}\label{16}
Q({\boldsymbol \alpha}) &= \sum_{l=1}^L {\rm E}\left[{\rm log}~{\mathcal N}\left(\bar{\mathbf x}_l; {\mathbf 0}, {\rm diag}({\boldsymbol \alpha}^{-1}(t))\right) \right]+ \sum_{i=1}^N {\rm log}~p(\alpha_i)+ {\rm const}  \notag\\
& = - \frac{1}{2}\sum_{l=1}^L{\rm E}\left[\bar{\mathbf x}_l^{\rm T}{\rm diag}({\boldsymbol \alpha}(t)) \bar{\mathbf x}_l\right]+ \frac{L}{2}\sum_{i=1}^N{\rm log}~{\alpha_i}+ \sum_{i=1}^N\left[(a - 1){\rm log}~{\alpha_i} - b{\alpha_i}\right]  + {\rm const}
\end{align}where the expectation is taken with respect to the posterior distribution of $\bar{\mathbf x}_l$, $\bar{\mathbf x}_l = [{\rm Re}({\mathbf x}_l^{\rm T} ), {\rm Im}({\mathbf x}_l^{\rm T} )]^{\rm T}$.
Performing the M step and setting $\partial {Q({\boldsymbol \alpha}) }/\partial{\boldsymbol \alpha}={\mathbf 0}$, we update ${\boldsymbol \alpha}(t+1)$ as
\begin{align}\label{alpha_update_mmv}
{\boldsymbol \alpha}(t+1)=\frac{L+2a-2}{2b+L{\rm diag}(\bar{\boldsymbol \Sigma}_{\rm post}(t))+\sum\limits_{l=1}^L(\bar{\boldsymbol \mu}_{{\rm post},l}(t))^2}.
\end{align}
The whole process is summarized in Algorithm \ref{algorithmmultiple}.
\begin{algorithm}[h]
\caption{BSBL algorithm for one-bit compressed sensing with MMVs}
\begin{algorithmic}[1]\label{algorithmmultiple}
\STATE Initialize ${\boldsymbol \alpha}(1)$, $\bar{\boldsymbol \mu}_{\rm post}(0)$ and $\bar{\boldsymbol \Sigma}_{\rm post}(0)$; Set the parameters of the noise covariance matrix {${\mathbf C}_{\bar{\mathbf w}}$}, the number of iterations $T$ and the damping factor $\gamma$;
\FOR {$t=1,\cdots,T$ }
\STATE For each snapshot, perform the E step and calculate the post means and variances of $\bar{\mathbf x}_l$ as (\ref{post_x}).
\STATE Perform the damping step (\ref{damp}).
\STATE Perform the M step and update ${\boldsymbol \alpha}$ as ${\boldsymbol \alpha}(t+1)$ (\ref{alpha_update_mmv}).
\ENDFOR
\STATE Return $\hat{\mathbf X}$.
\end{algorithmic}
\end{algorithm}

\section{Simulation}
In this section, we compare the performance of the BSBL algorithm against SVM \cite{yulong}, BIHT and Gr-SBL methods to solve the DOA estimation problem, where there exist three signal sources at direction of arrivals (DOAs) $[-3, 2, 75]^{\circ}$ with amplitudes being $[12, 22, 20]$ dB in the uniform linear array (ULA) scenario, {which is the same scenario as used in \cite{PeterSPL}}. The relationship between the interelement spacing $d$ and the wavelength $\lambda$ is $d = \lambda/2$. All the DOAs are restricted into an angular grid $[-90: 0.5: 90]^{\circ}$.  Each column of the measurement matrix $\mathbf A$ is defined as ${\mathbf a}(\theta_l) \triangleq \frac{1}{\sqrt{M}}\left[1,e^{-{\rm j}2\pi d{\rm {sin}}{\theta}_l/\lambda},\cdots, e^{-{\rm j}2\pi (M-1)d{\rm {sin}}{\theta}_l/\lambda}\right]^{\rm T}.$ In both experiments, the signal-to-noise ratio (SNR) in SMV and MMVs is defined as ${\rm SNR}_{\rm s}=10\log_{10}\frac{{\rm E}\|{\mathbf A}{\mathbf x}\|_2^2}{{\rm E}\|{\mathbf w}\|_2^2}=10\log_{10}\frac{{\rm E}\|{\mathbf A}{\mathbf x}\|_2^2}{M\sigma_w^2}$ and  ${\rm SNR}_{\rm m}=10\log_{10}\frac{{\rm E}\|{\mathbf A}{\mathbf X}\|_{\rm F}^2}{{\rm E}\|{\mathbf W}\|_{\rm F}^2}=10\log_{10}\frac{{\rm E}\|{\mathbf A}{\mathbf X}\|_{\rm F}^2}{ML\sigma_w^2}$, respectively, {where $\|\cdot\|_2$ and $\|\cdot\|_{\rm F}$ denotes the $l_2$ and Frobenius norm separately}. Then we calculate the noise variance $\sigma_w^2$ according to the $\rm SNR$. For both the Gr-SBL and BSBL algorithm, we set $a=1$, $b=0$ which corresponds to the uninformative prior of $\boldsymbol \alpha$. We set the number of iterations as 500 for both BSBL and Gr-SBL methods. The damping factor is $\gamma = 0.6$.

\begin{figure}[h!t]
\centering{\includegraphics[width=85mm]{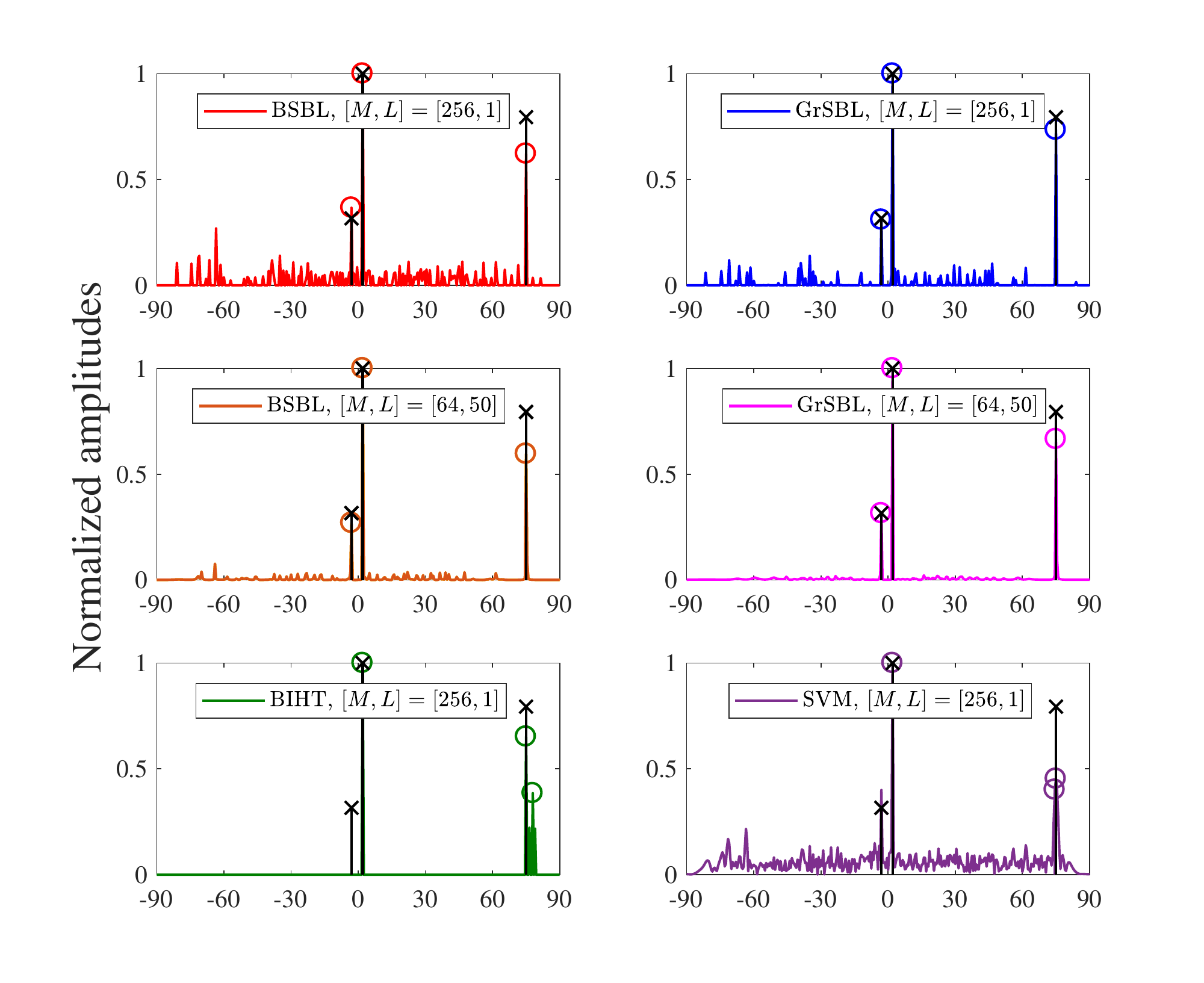}}
\caption{Estimation performance of various algorithms at ${\rm SNR} = 10$ dB. The circle markers denote the top $K$ magnitudes, and the cross markers denote the true DOAs.\label{mu}}
\end{figure}

Before moving forward, we firstly conduct a experiment about estimation performance of various algorithms in one single realization. The results are presented in Fig. \ref{mu}. Given that the number of sources is known, it can be seen that all algorithms except BIHT and SVM successfully locate the DOAs of the three sources. In the SMV scenarios, the Gr-SBL has the best estimation performance. As for the MMVs scenario, the performance of the BSBL algorithm is comparable to that of the Gr-SBL algorithm. In addition, we can see that increasing the number of snapshots is very beneficial for enhancing the reconstruction performance.

In the following, we conduct the two numerical experiments when the noise is uncorrelated. We have found that the proposed method also works well with noise being correlated.

In the first simulation, we want to compare the estimation performance of all algorithms in terms of normalized mean square error (NMSE). The debiased NMSE for SMV and MMVs scenarios is defined as $\underset{ c}{\operatorname{min}}~10\log ({\|{\mathbf x}-c\hat{\mathbf x}\|_2}/{\|{\mathbf x}\|_2})$ or $\underset{\mathbf c}{\operatorname{min}}~10\log ({\|{\mathbf X}-{\rm diag}({\mathbf c})\hat{\mathbf X}\|_{\rm F}})/{\|{\mathbf X}\|_{\rm F}}$, respectively. For all four algorithms, we look into the relationship between the debiased NMSE of the algorithms against the SNR in the scenario where $M = 256$, $L = 1$ and $M = 64$, $L = 50$, respectively. For both the Gr-SBL and BSBL algorithm, the mismatched corresponds to the scenario that we set the input noise variance to 1 instead of the true one.
\begin{figure}[h!t]
\centering{\includegraphics[width=85mm]{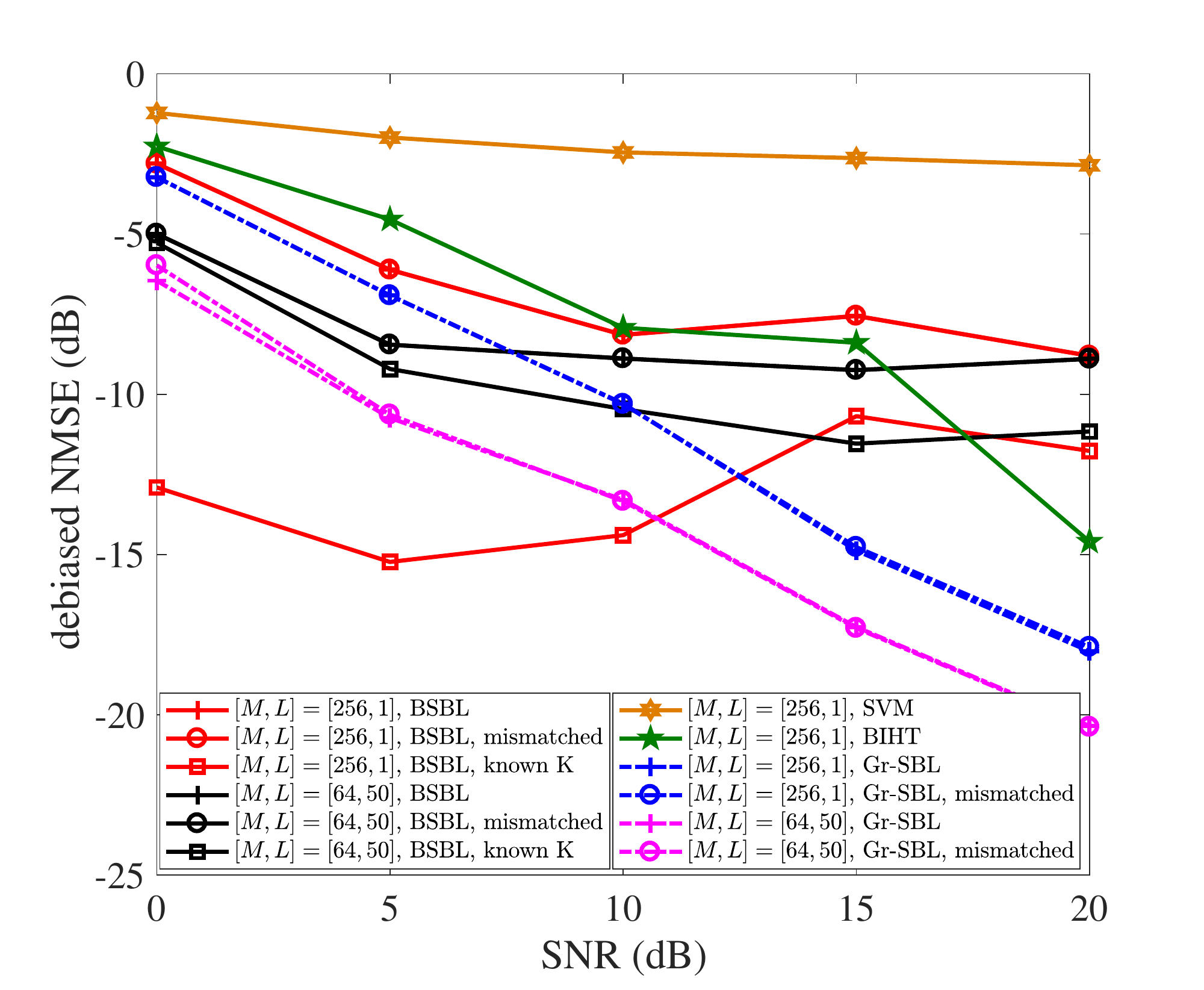}}
\caption{The debiased NMSE versus SNR averaged over 10 Monte Carlo (MC) trials.\label{NMSE1}}
\end{figure}

It can be seen from Fig. \ref{NMSE1} that the performance of the BSBL and Gr-SBL methods in the mismatched scenario is the same as that of the variance-known cases. The debiased NMSE of the BSBL is larger than that of BIHT for ${\rm SNR}\geq 15{\rm dB}$. The reason is that the the number of the unknown sources $K$ is available to the BIHT method. Actually, if the BSBL chooses the top $K$ magnitudes, it achieves a lower debiased NMSE, as seen in Fig. \ref{NMSE1}. Utilizing the information of $K$ significantly reduces the debiased NMSE of the BSBL algorithm in the SMV settings. In contrast to the SMV settings, the effect of utilizing $K$ becomes smaller in the MMVs settings. The reason is that the BSBL algorithm without knowing $K$ can locate the source more accurately in the MMVs settings, as shown in Fig. \ref{mu}. {As for the running time, from Table \ref{table1}, we can see that the BIHT is the fastest, and the Gr-SBL with MMVs is the slowest. The running time of the BSBL with MMVs is much faster than that of itself in the SMV setting.}
\begin{figure}[h!t]
\centering{\includegraphics[width=85mm]{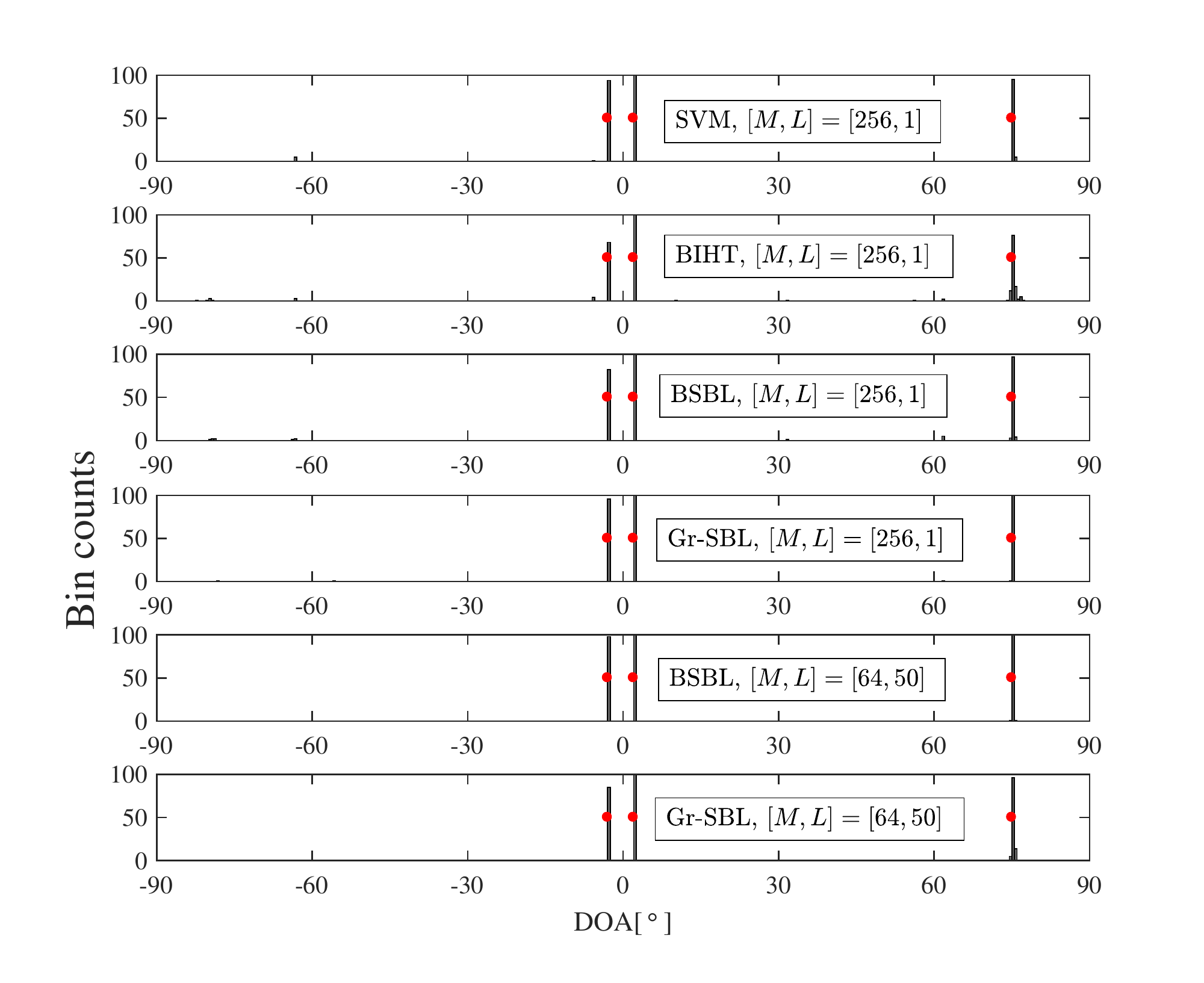}}
\caption{Bin counts of the SVM, BIHT, Gr-SBL and BSBL based on 100 MC trials.\label{BIN}}
\end{figure}
\begin{table}[h!t]
    \begin{center}
\caption{{Running time (seconds) of algorithms averaged over 100 MC trials.}}\label{table1}
    \begin{tabular}{|c|c|c|c|c|c|c|}
            \hline
            SVM & BSBL (L = 1) & Gr-SBL (L = 1)\\ \hline
            8.3 &    69.8 & 31.3 \\ \hline
          BIHT &  BSBL (L = 50) &Gr-SBL (L = 50)\\ \hline
            0.8 & 15.5 & 209.4\\ \hline
             \end{tabular}
    \end{center}
\end{table}
\begin{table}[h!t]
    \begin{center}
\caption{{Number of successively detecting true DOAs in 100 MC trials.}}\label{table2}
    \begin{tabular}{|c|c|c|c|c|c|c|}
            \hline
            SVM & BSBL (L = 1) & Gr-SBL (L = 1)\\ \hline
            89&    79 & 96 \\ \hline
          BIHT &  BSBL (L = 50) &Gr-SBL (L = 50)\\ \hline
            47 & 98&81\\ \hline
             \end{tabular}
    \end{center}
\end{table}

As for the second experiment, we plot the bin counts over 100 MC simulations at ${\rm SNR} = 10$ dB. {From Fig. \ref{BIN} and Table \ref{table2}, we can see that the Gr-SBL algorithm has the best location results and the BIHT is the worst in the SMV scenario.} For the MMVs scenarios, the BSBL algorithm has a better estimation performance than Gr-SBL algorithm. In general, the BSBL method with MMVs achieves the best location performance.

\section{Conclusion}
{We propose a BSBL algorithm to cope with the one-bit CS reconstruction problems. The proposed algorithm transfers the original nonlinear problem into a linear one and naturally takes the standard SBL framework into consideration. Algorithms for both the SMV and MMVs scenarios are introduced. Furthermore, simulations demonstrate the effectiveness of the proposed algorithm, especially in the MMVs scenario.}

\newpage{}
\bibliographystyle{IEEEbib}
\bibliography{strings,refs}

\begin{thebibliography}{}
\bibitem{Donoho1}
 D. L. Donoho, ``Compressed sensing,'' \emph{IEEE Trans. Inf. Theory}, vol. 52, no. 4, pp. 1289-1306, 2006.
\bibitem{Donoho2}
M. Lustig, D. L. Donoho, J. M. Santos and J. M. Pauly, ``Compressed sensing MRI,'' \emph{IEEE Signal Processing Magazine}, vol. 25, no. 2, pp. 72-82, 2008.
\bibitem{Huang}
G. Huang, H. Jiang, K. Matthews and P. Wilford, ``Lensless imaging by compressive sensing,'' \emph{2013 IEEE International Conference on Image Processing}, pp. 2101-2105, 2013.
\bibitem{Firooz}
M. H. Firooz and S. Roy, ``Network tomography via compressed sensing,'' \emph{2010 IEEE Global Telecommunications Conference GLOBECOM}, pp. 1-5, 2010.

\bibitem{Robert}
R. Tibshirani, ``Regression shrinkage and selection via the Lasso,'' \emph{Journal of the Royal Statistical Society, Series B (Methodological)}, vol. 58, no. 1, pp. 267-288. JSTOR, 1996.
\bibitem{Daubechies}
I. Daubechies, M. Defrise and C. D. Mol, ``An iterative thresholding algorithm for linear inverse problems with a sparsity constraint,'' \emph{Commun. Pure Appl. Math.}, vol. 57, pp. 1413-1457, 2004.

\bibitem{Tipping}
M. E. Tipping, ``Sparse Bayesian learning and the relevance vector
machine,'' \emph{JMLR}, vol. 1, pp. 211-244, 2001.
\bibitem{WipfS}
D. P. Wipf and B. D. Rao, ``Sparse Bayesian learning for basis selection,'' \emph{IEEE Trans. Signal Process.}, vol. 52, no. 8, pp. 2153-2164, 2004.

\bibitem{Donoho}
 D. L. Donoho, A. Maleki and A. Montanari, ``Message passing algorithms for compressed sensing,'' \emph{PNAS,} vol. 106, no. 45, pp. 18914-18919, 2009.

\bibitem{Singh}
J. Singh, O. Dabeer and U. Madhow, ``On the limits of communication with low-precision analog-to-digital conversion at the receiver,'' \emph{IEEE Trans. Commun.,} vol. 57, no. 12, pp. 3629-3639, 2009.
\bibitem{BIHT}
L. Jacques, J. L. Laska, P. T. Boufounos and R. G. Baraniuk, ``Robust 1-bit compressive sensing via binary stable embeddings of sparse vectors,'' \emph{IEEE Trans. Inf. Theory}, pp. 2082-2102, 2013.

\bibitem{Rangan1}
S. Rangan, ``Generalized approximate message passing for estimation with random linear mixing,'' in \emph{Proc. IEEE Int. Symp. Inf. Theory}, pp. 2168-2172, 2011.
\bibitem{Mo}
J. Mo, P. Schniter, N. G. Prelcic and R. W. Heath Jr, ``Channel estimation in millimeter wave MIMO systems with one-bit quantization,'' \emph{Proc. IEEE Asilomar Conf. Signals Syst. Comput.,} pp. 957-961, 2014.
\bibitem{Hangting}
H. Cao, J. Zhu and Z. Xu, ``Adaptive one-bit quantization via approximate message passing with nearest neighbour sparsity pattern learning,'' \emph{IET Signal Processing}, 2018.
\bibitem{Kamilov}
U. S. Kamilov, A. Bourquard, A. Amini and M. Unser, ``One-bit measurements with adaptive thresholds'', \emph{IEEE Signal Process. Lett.}, vol. 19, no. 10, pp. 607-610, 2012.



\bibitem{meng1}
X. Meng, S. Wu and J. Zhu, ``A unified Bayesian inference framework for generalized linear models,'' \emph{IEEE Signal Process. Lett.}, vol. 25, no. 3, pp. 398-402, 2018.
\bibitem{meng2}
X. Meng and J. Zhu, ``A generalized sparse Bayesian learning algorithm for one-bit DOA estimation,'' to appear in \emph{IEEE Commun. Lett.}, 2018.


\bibitem{Bussgang}
J. Bussgang, ``Cross-correlation function of amplitude-distorted gaussian signals,'' MIT, Res. Lab. Elec. Tech. Rep. 216, Mar. 1952.
\bibitem{Andrew}
A. S. Lan, M. Chiang and C. Studer, ``Linearized binary regression,'' available at https://arxiv.org/pdf/1802.00430.pdf.
\bibitem{SBL}
K. P. Murphy, \emph{Machine Learning: a probabilistic perspective}, pp. 463-467, 2012.
\bibitem{Li}
Y. Li, C. Tao, G. Seco-Granados, A. Mezghani, A. Swindlehurst and L. Liu, ``Channel estimation and performance analysis of one-bit massive MIMO systems,'' \emph{IEEE Trans. Signal Process.,} vol. 65, no. 15, pp. 4075-4089, 2017.

\bibitem{yulong}
Y. Gao, D. Hu, Y. Chen and Y. Ma, ``Gridless 1-b DOA estimation exploiting SVM approach,'' \emph{IEEE Commun. Lett.}, vol. 21, no. 4, Oct. 2017.
\bibitem{PeterSPL}
P. Gerstoft, C. F. Mecklenbr\"{a}uker, A. Xenaki and S. Nannuru, ``Multisnapshot sparse Bayesian learning for DOA,'' \emph{IEEE Signal Process. Lett.}, vol. 23, no. 10, pp. 1469-1473, 2016.

\end{thebibliography}

\end{document}